\documentclass[12pt]{article}
\usepackage{epsfig,graphicx}
\usepackage{fpcp03}

\def\beq{\begin{equation}}
\def\eeq#1{\label{#1}\end{equation}}
\def\eeqn{\end{equation}}

\def\beqa{\begin{eqnarray}}
\def\eeqa#1{\label{#1}\end{eqnarray}}
\def\eeqan{\end{eqnarray}}

\let\bar=\overbar

\def\Dslash{\not{\hbox{\kern-4pt $D$}}}
\def\dslash{\not{\hbox{\kern-2pt $\del$}}}

\def\msb{{\bar{\ssstyle M \kern -1pt S}}}

\def\BB0bar{B^0 {\overline B}^0}
\def\BB0dbar{B_d^0 {\overline B}_d^0}
\def\BB0sbar{B_s^0 {\overline B}_s^0}

\RequirePackage{xspace}

\usepackage{relsize}
\def\babar{\mbox{\slshape B\kern-0.1em{\smaller A}\kern-0.1em
    B\kern-0.1em{\smaller A\kern-0.2em R}}}

\def\Kbar  {\kern 0.2em\overline{\kern -0.2em K}{}\xspace}

\def\Kz    {\ensuremath{K^0}\xspace}
\def\Kzb   {\ensuremath{\Kbar^0}\xspace}
\def\KzKzb {\ensuremath{\Kz \kern -0.16em \Kzb}\xspace}
\def\Kp    {\ensuremath{K^+}\xspace}
\def\Km    {\ensuremath{K^-}\xspace}

\def\KpKm  {\ensuremath{\Kp \kern -0.16em \Km}\xspace}

\def\Dbar    {\kern 0.2em\overline{\kern -0.2em D}{}\xspace}

\def\Dz      {\ensuremath{D^0}\xspace}
\def\Dzb     {\ensuremath{\Dbar^0}\xspace}
\def\DzDzb   {\ensuremath{\Dz {\kern -0.16em \Dzb}}\xspace}
\def\Dp      {\ensuremath{D^+}\xspace}
\def\Dm      {\ensuremath{D^-}\xspace}

\def\DpDm    {\ensuremath{\Dp {\kern -0.16em \Dm}}\xspace}

\def\Bbar    {\kern 0.18em\overline{\kern -0.18em B}{}\xspace}

\def\BB      {\ensuremath{B\Bbar}\xspace} 
\def\Bz      {\ensuremath{B^0}\xspace}
\def\Bzb     {\ensuremath{\Bbar^0}\xspace}
\def\BzBzb   {\ensuremath{\Bz {\kern -0.16em \Bzb}}\xspace}
\def\Bu      {\ensuremath{B^+}\xspace}
\def\Bub     {\ensuremath{B^-}\xspace}

\def\BpBm    {\ensuremath{\Bu {\kern -0.16em \Bub}}\xspace}

\mathchardef\Upsilon="7107
\def\Y#1S{\ensuremath{\Upsilon{(#1S)}}\xspace}% no space before {...}!

\mathchardef\Deltares="7101
\mathchardef\Xi="7104
\mathchardef\Lambda="7103
\mathchardef\Sigma="7106
\mathchardef\Omega="710A

\def\Deltabar{\kern 0.25em\overline{\kern -0.25em \Deltares}{}\xspace}
\def\Lbar{\kern 0.2em\overline{\kern -0.2em\Lambda\kern 0.05em}\kern-0.05em{}\xspace}
\def\Sigbar{\kern 0.2em\overline{\kern -0.2em \Sigma}{}\xspace}
\def\Xibar{\kern 0.2em\overline{\kern -0.2em \Xi}{}\xspace}
\def\Obar{\kern 0.2em\overline{\kern -0.2em \Omega}{}\xspace}
\def\Nbar{\kern 0.2em\overline{\kern -0.2em N}{}\xspace}
\def\Xb{\kern 0.2em\overline{\kern -0.2em X}{}\xspace}

\newcommand{\tev}{\ensuremath{\mathrm{\,Te\kern -0.1em V}}\xspace}
\newcommand{\gev}{\ensuremath{\mathrm{\,Ge\kern -0.1em V}}\xspace}
\newcommand{\mev}{\ensuremath{\mathrm{\,Me\kern -0.1em V}}\xspace}
\newcommand{\kev}{\ensuremath{\mathrm{\,ke\kern -0.1em V}}\xspace}
\newcommand{\ev}{\ensuremath{\mathrm{\,e\kern -0.1em V}}\xspace}
\newcommand{\gevc}{\ensuremath{{\mathrm{\,Ge\kern -0.1em V\!/}c}}\xspace}
\newcommand{\mevc}{\ensuremath{{\mathrm{\,Me\kern -0.1em V\!/}c}}\xspace}
\newcommand{\gevcc}{\ensuremath{{\mathrm{\,Ge\kern -0.1em V\!/}c^2}}\xspace}
\newcommand{\mevcc}{\ensuremath{{\mathrm{\,Me\kern -0.1em V\!/}c^2}}\xspace}
\def\mus  {\ensuremath{\rm \,\mus}\xspace}

\def\mus        {\ensuremath{\,\mu{\rm s}}\xspace}    %% microsecond
\def\pep2{PEP-II}

\def\gsim{{~\raise.15em\hbox{$>$}\kern-.85em
          \lower.35em\hbox{$\sim$}~}\xspace}
\def\lsim{{~\raise.15em\hbox{$<$}\kern-.85em
          \lower.35em\hbox{$\sim$}~}\xspace}

\xspace

\def\jetset74   {\mbox{\tt Jetset \hspace{-0.5em}7.\hspace{-0.2em}4}\xspace}
\begin{document}

%+ \Chapter{}
%+ {Instruction for producing FPCP 2003 proceedings}
%+ {Pascal~Perret}

\Title{Cosmology and CP Violation \footnote{Based on work done 
in collaboration with R. Gonz\' alez Felipe, F. R. Joaquim, I.
Masina, M. N. Rebelo and C. A. Savoy  in \cite{varios} }}
\bigskip

%+ \addcontentsline{toc}{chapter}{{\it Pascal~Perret}}
%+ \index{author}{Perret, P.} 

%%%%%%%%%%%%%%%%%%%%%%%%%%%%%%%%%%%%%
% Label to flag the first page of your contribution
% Replace Perret by your name starting with a capital letter
%
\label{PerretStart}

%%%%%%%%%%%%%%%%%%%%%%%%%%%%%%%%%%%%%
% Your name
%
\author{ Gustavo C.Branco\index{Branco, G.C.} }

%%%%%%%%%%%%%%%%%%%%%%%%%%%%%%%%%%%%%
% Your address
%
\address{Departamento de Fisica and Grupo Teorico de Fisica de Particulas \\
Instituto Superior Tecnico\\
Av. Rovisco Pais,1049-001 Lisboa, Portugal\\ 
\vspace{1.0cm}
Invited Talk at  Flavor Physics And CP Violation (FPCP 2003) \\
3-6 Jun 2003, Paris, France
}

\makeauthor\abstracts{ 
We describe the role of CP violation in the generation of the 
baryon asymmetry of the Universe, in the framework of baryogenesis through leptogenesis,
with emphasis on the possible relationship between CP violation at low energies and that 
required by leptogenesis. It is emphasized that a direct link between these two 
manifestations of CP violation only exists in the framework of specific flavour structures for
the fundamental leptonic mass matrices.}

\section{Introduction}

The phenomenon of CP violation has profound implications for Cosmology, since it 
is one of the necessary ingredients \cite{Sakharov:dj}
for generating the observed baryon 
asymmetry of the Universe (BAU).
During the last few years, the data collected from the acoustic peaks in the
cosmic microwave background radiation~\cite{Jungman:1995bz} has allowed to
obtain a precise measurement of BAU. The
MAP experiment~\cite{MAP} and the PLANCK satellite~\cite{PLANCK} planned for
the near future should further improve this result. At the present time, the
measurement of the baryon-to-entropy ratio $Y_B=n_B/s$ is
\begin{equation}
\label{YBrng}%
0.7 \times 10^{-10} \lsim Y_B\lsim 1.0 \times 10^{-10}\,.
\end{equation}

A great challenge for Particle Physics is finding a plausible mechanism
capable of reproducing this ratio.
Although there are various possible scenarios for baryogenesis
one of the most appealing ones is that provided by 
leptogenesis \cite{Fukugita:1986hr}, 
where first the out-of-equilibrium decay of righthanded neutrinos
creates a lepton asymmetry which is then converted into
a baryon asymmetry through B-violating but (B-L) conserving
sphaleron mediated processes \cite{Kuzmin:1985mm}.
In any baryogenesis scenario, a fascinating question which naturally 
arises is whether low energy data on CP violation obtained from terrestrial experiments
could give us information on the processes responsible for the creation of BAU. 
More specifically, in the context of leptogenesis, one may wonder 
whether it is possible
to relate CP violation necessary to generate BAU, to leptonic 
CP violation at low energies \cite{varios}, 
\cite{Branco:2001pq},  \cite{Rebelo:2002wj}, observable through neutrino oscillations. 
It has been shown 
that this connection exists only in 
specific models \cite{Branco:2001pq}, \cite{Rebelo:2002wj}. \\

\section{A Minimal Extension of the Standard Model}
In order to understand the relationship between CP breaking at low energies and
CP violation responsible for leptogenesis, one has to specify the particle physics 
framework one is considering.  
We will work within a minimal extension of the Standard Model (SM) which consists of
adding to the standard spectrum, one right-handed neutrino per generation.
At this stage, no other assumption is beeing made, so our framework is quite
general and it is indeed the simplest extension of the SM capable of generating
non vanishing but naturally small neutrino masses.
Before gauge symmetry breaking, the leptonic couplings to the SM Higgs doublet
$\phi$ can be written as:
\begin{equation}
{\cal L}_Y  = - Y_\nu\left(\bar{\ell}_L^{\;0},
\bar{\nu}_L^{\;0}\right)\widetilde {\phi}\,\nu_{R}^{\,0} -
Y_{\ell} \left(\bar{\ell}_L^{\;0}, \bar{\nu}_L^{\;0}\right)\phi \
\ell_{R}^{\,0} + {\rm H.c.}\,, \label{Lyuk}
\end{equation}
where $\widetilde {\phi} = i \tau _2 \phi ^\ast $. After
spontaneous gauge symmetry breaking, the leptonic mass terms are
given by:
\begin{eqnarray}
{\cal L}_m  &= -\left[\,\bar{\nu}_L^{\;0} m_D \nu_{R}^{\,0} + \frac{1}{2}
\nu_{R}^{0\,T} C M_R \nu_{R}^{\,0}+
\bar{\ell}_L^{\;0} m_{\ell}\,\ell_R^{\,0}\,\right] + {\rm H.c.} \nonumber \\
&= - \left[\frac{1}{2}\,n_{L}^{T} C {\cal M}^* n_L+
\bar{\ell}_L^{\;0} m_{\ell}\,\ell_R^{\,0} \right] + {\rm H.c.}
\,,\label{Lmass}
\end{eqnarray}
where $m_D = v\,Y_\nu$ is the Dirac neutrino  mass matrix with $v=\langle
\phi^{\,0} \rangle /\sqrt{2} \simeq 174\,$GeV, $M_R$ and $m_\ell=v\,Y_{\ell}$
denote the right-handed Majorana neutrino and charged lepton mass matrices,
respectively, and $n_L= ({\nu}_{L}^0, {(\nu_R^0)}^c)$. Among all the terms,
only the right-handed neutrino Majorana mass term is SU(2) $\times$ U(1)
invariant and, as a result, the typical scale of $M_R$ can be much above  the
electroweak symmetry breaking scale $v$, thus leading to naturally small
left-handed Majorana neutrino masses of the order $m^2_D/M_R$ through the
seesaw mechanism. In terms of weak-basis eigenstates the leptonic charged
current interactions are given by:
\begin{equation}
{\cal L}_W  = -\frac{g}{\sqrt 2}W^{-}_{\mu}\,\bar{\ell}_L^{\;0} \,
\gamma ^{\mu}\,\nu_L^{\,0} + {\rm H.c.}\,. \label{Lcc1}
\end{equation}
It is clear from Eqs. (\ref{Lmass}) and (\ref{Lcc1}) that it is possible to
choose, without loss of generality, a weak basis (WB) where both $m_\ell$ and
$M_R$ are diagonal, real and positive. Note that in this WB, $m_D$ is a general
complex matrix which contains all the information on CP-violating phases
as well as on leptonic mixing.
Since we are considering a standard Higgs sector,
in the present framework there is no $\Delta L=2$ mass term of the form
$\frac{1}{2} \nu_{L}^{0\,T} C M_L \nu_{L}^0$ at tree level.
The total number of CP-violating
phases for $n$ generations is then given by $n(n -1)$ \cite{Endoh:2000hc} 
since one can eliminate $n$ of the initial $n^2$ phases of $m_D$.
All CP violating phases are 
contained in $m_D$ in this special weak basis\footnote{The counting of
independent CP-violating phases for the general case, where besides $m_D$ and
$M_R$ there is also a left-handed Majorana mass term at tree level has been
discussed in Ref.~\cite{Branco:gr}.}.

In the physical basis (i.e. the mass eigenstates basis ) all CP violating phases 
are shifted to the leptonic mixing matrix appearing in charged weak currents.
We recall that the full $6 \times 6$ neutrino mass matrix $\mathcal{M}$ is
diagonalized via the transformation:
\begin{equation}
V^T {\cal M}^* V = \cal D , \label{Mnudi}
\end{equation}
where ${\cal D} ={\rm diag} (m_1, m_2, m_3, M_1, M_2, M_3)$, with $m_i$ and
$M_i$ denoting the physical masses of the light and heavy Majorana neutrinos,
respectively. It is convenient to write $V$ and $\cal D$ in the following form,
together with the definition of $\cal M$ :
\begin{equation}
V= \left (\begin{array}{cc}
K & Q \\
S & T \end{array}\right) \;\;,\;\;{\cal D}=\left(\begin{array}{cc}
d_\nu & 0 \\
0 & D_R \end{array}\right)\;\;,\;\;{\cal M}= \left (\begin{array}{cc}
0 & m_D \\
m_D^{\,T} & M_R \end{array}\right).
\end{equation}
From Eq. (\ref{Mnudi}) one obtains, to an excellent approximation, the seesaw
formula:
\begin{equation}
d_\nu\simeq -K^\dagger\, m_D\, M_R^{-1}\, m_D^{\,T}\, K^* \equiv
K^\dagger\,\mathcal{M}_\nu\, K^*\,, \label{ssaw}
\end{equation}
where $\mathcal{M}_\nu$ is the usual light neutrino effective mass
matrix. The leptonic charged-current interactions are given by:
\begin{equation}
- \frac{g}{\sqrt{2}} \left( \bar{\ell}_{L} \,\gamma_{\mu} \, K
{\nu}_{L} + \bar{\ell}_{L}\, \gamma_{\mu}\, Q\, N_{L} \right)
W^{\mu} +{\rm H.c.}\,, \label{Lcc2}
\end{equation}
where $\nu_i$ and  $N_i$ denote the light and heavy neutrino mass eigenstates,
respectively. The matrix $K$ which contains all information on mixing and CP
violation at low energies can then be parametrized, after eliminating the
unphysical phases, by $K= { U_{\delta}} P$ with $P ={\rm diag}(1,
e^{i\,\alpha},e^{i\,\beta})$ ($\alpha$ and $\beta$ are Majorana phases) and
$U_{\delta}$ a unitary matrix which contains only one (Dirac-type) phase
$\delta$. In the limit where the heavy neutrinos exactly decouple from the
theory, the matrix $K$ is usually referred as the
Pontecorvo-Maki-Nakagawa-Sakata mixing matrix, which from now on we shall
denote as $U_\nu$. It is clear from Eq.(\ref{ssaw}) that in general and without 
further assumptions the phases $\alpha$ , $\beta$ and $\delta$ are complicated functions 
of the six independent phases appearing in $m_D$ ( we are considering $3$ generations )
in the weak-basis where $m_l$ and $M_R$ are diagonal, real and positive. As we will see 
in the sequel, this is the essential reason why in general and without further assumptions,
it is not possible to establish 
a direct connection between the phases appearing at low energies and those relevant 
for leptogenesis.

\section {CP Violation in Neutrino Oscillations}

It has been shown \cite{Branco:2001pq} that the strength of CP violation at low
energies, observable for example through neutrino oscillations, can be obtained
from the following low-energy WB invariant:
\begin{eqnarray}
{\cal T}_{CP} = {\rm Tr}\left[\,\mathcal{H}_{\nu}, H_\ell
\,\right]^3=6\,i \,\Delta_{21}\,\Delta_{32}\,\Delta_{31}\,{\rm Im}
\left[\,
(\mathcal{H}_{\nu})_{12}(\mathcal{H}_{\nu})_{23}(\mathcal{H}_{\nu})_{31}\,
\right]\,, \label{TCP}
\end{eqnarray}
where $\mathcal{H}_{\nu}=\mathcal{M}_\nu\,\mathcal{M}_\nu^{\dag}$,
$H_\ell=m_\ell\,{m_\ell}^{\dagger}$ and $\Delta_{21}=({m_{\mu}}^2-{m_e}^2)$
with analogous expressions for $\Delta_{31}$, $\Delta_{32}$. This relation can
be computed in any weak basis. This is specially useful since most of the ansatze 
for the leptonic mass matrices are written in a WB where neither $\mathcal{H}_{\nu}$
nor $H_\ell$ are diagonal. The above WB invariant enables one to investigate whether
a specific ansatz leads to CP violation in neutrino oscillations or not, without
performing any diagonalization of leptonic mass matrices and without computing $U_\nu$.  
The low-energy invariant (\ref{TCP}) is
sensitive to the Dirac-type phase $\delta$ and vanishes for $\delta=0$. On the
other hand, it does not depend on the Majorana phases $\alpha$ and $\beta$
appearing in the leptonic mixing matrix. The quantity ${\cal T}_{CP}$ can be
fully written in terms of physical observables since
\begin{eqnarray}
{\rm Im} \left[\, (\mathcal{H}_{\nu})_{12}(\mathcal{H}_{\nu})_{23}
(\mathcal{H}_{\nu})_{31}\, \right] = - \Delta m^2_{21} \,
\Delta m^2_{31}  \, \Delta m^2_{32}  {\cal
J}_{CP}\,, \label{ImHHH}
\end{eqnarray}
where the $\Delta m_{ij}^2$'s are the light neutrino mass squared
differences and ${\cal J}_{CP}$ is the imaginary part of an invariant quartet
appearing in the difference of the CP-conjugated neutrino oscillation
probabilities $P(\nu_e\rightarrow\nu_\mu)-P(\bar{\nu}_e\rightarrow
\bar{\nu}_\mu)$. One can easily get:
\begin{eqnarray}
{\cal J}_{CP} &\equiv {\rm Im}\left[\,(U_\nu)_{11} (U_\nu)_{22}
(U_\nu)_{12}^\ast (U_\nu)_{21}^\ast\,\right] \nonumber \\
&= \frac{1}{8} \sin(2\,\theta_{12}) \sin(2\,\theta_{13}) \sin(2\,\theta_{23})
\cos(\theta_{13})\sin \delta\,, \label{Jgen1}
\end{eqnarray}
where the $\theta_{ij}$ are the mixing angles appearing in the standard
parametrization adopted in \cite{Hagiwara:fs}. Alternatively, one can use
Eq.~(\ref{ImHHH}) and write:
\begin{eqnarray} {\cal J}_{CP}=-\frac{{\rm Im}\left[\,
(\mathcal{H}_{\nu})_{12}(\mathcal{H}_{\nu})_{23}(\mathcal{H}_{\nu})_{31}\,
\right]}{\Delta m^2_{21}  \, \Delta m^2_{31}
\,\Delta m^2_{32}  }\,. \label{Jfin}
\end{eqnarray}
This expression  again has the advantage of allowing the computation of the leptonic low-energy
CP rephasing invariant ${\cal J}_{CP}$, without resorting to the mixing matrix $U_\nu$.

It is also possible to write WB invariants useful to leptogenesis
\cite{Branco:2001pq} as well as WB invariant conditions for CP conservation in
the leptonic sector relevant in specific frameworks
\cite{Branco:gr}, \cite{Branco:1999bw}.

\section{\bf CP Asymmetries in Heavy Majorana Neutrino Decays}
\label{CPasymmetries}%
The starting point in leptogenesis scenarios is the $CP$ asymmetry generated
through the interference between tree-level and one-loop heavy Majorana
neutrino decay diagrams. In the simplest extension of the SM, such diagrams
correspond to the decay of the Majorana neutrino into a lepton and a Higgs
boson. Considering the decay of one heavy Majorana neutrino $N_j$, this
asymmetry is given by:
\begin{eqnarray} \label{epsin}
\varepsilon_j=\frac{\Gamma\,(N_j \rightarrow \ell\,\phi)-\Gamma
\,(N_j \rightarrow \bar{\ell}\,\phi^{\,\dag})}{\Gamma\,(N_j
\rightarrow \ell\,\phi)+\Gamma\,(N_j \rightarrow
\bar{\ell}\,\phi^{\,\dag})}\ .
\end{eqnarray}
In terms of the Dirac neutrino Yukawa couplings the CP asymmetry (\ref{epsin})
is \cite{Covi:1996wh}:
\begin{eqnarray}
\label{epsj1} \varepsilon_j=\frac{1}{8\pi(Y_\nu^{\dag}
Y_\nu^{})_{jj}} \sum_{k\neq j}\,{\rm Im}
[\,(Y_\nu^{\dag}\,Y_\nu^{})_
{jk}^2\,]\,f\!\left (\frac{M_k^{\,2}}{M_j^{\,2}}\right)\,,
\end{eqnarray}
where the index $j$ is not summed over in $(Y_\nu^{\dag} Y_\nu^{})_{jj}\,$. The
loop function $f(x)$ includes the one-loop vertex and self-energy corrections
to the heavy neutrino decay amplitudes,
\begin{eqnarray}
\label{f} f(x)=\sqrt{x} \left[\,(1+x)\ln \left(\frac{x}
{1+x}\right)+\frac{2-x}{1-x}\,\right]\,.
\end{eqnarray}
From Eq.~(\ref{epsj1}) it can be readily seen that the CP asymmetries are only
sensitive to the CP-violating phases appearing in $Y_\nu^\dagger Y_\nu^{}$ (or
equivalently in $m_D^\dagger m_D^{}$) in the WB where $M_R$ and $m_\ell$ are
diagonal.

Let us consider
the hierarchical case $M_1 < M_2 \ll M_3$. In this case only the decay of the lightest
heavy neutrino $N_1$ is relevant for leptogenesis, provided the interactions of
$N_1$ are in thermal equilibrium at the time $N_{2,3}$ decay, so that the
asymmetries produced by the latter are erased before $N_1$ decays. In this
situation, it is sufficient to take into account the CP asymmetry
$\varepsilon_1$. Since in the limit $x \gg 1$ the function $f(x)$ can be
approximated by\footnote{This approximation can be reasonably used for $x
\gsim 15$.} $f(x)\simeq -3/(2\sqrt{x})$, we have from Eq.~(\ref{epsj1})
\begin{eqnarray}
\label{eps1} \varepsilon_1=-\frac{3}{16\pi(Y_\nu^{\dag}
Y_\nu^{})_{11}} \sum_{k=2,3}\,{\rm Im}
[\,(Y_\nu^{\dag}\,Y_\nu^{})_
{1k}^2\,]\,\frac{M_1}{M_k}\,,
\end{eqnarray}
which can be recast in the form \cite{Buchmuller:2000nd}
\begin{eqnarray}
\varepsilon_1 \simeq -\frac{3\,M_1}{16\,\pi}\frac{ {\rm
Im}\left[\,Y_\nu^\dagger\,Y_\nu^{}\,D_R^{-1}\,Y_\nu^T\,Y_\nu^\ast\,\right]
_{11}}{(Y_\nu^{\dag}Y_\nu^{})_{11}}=\frac{3\,M_1}{16\,\pi\,v^2}\frac{
{\rm Im}\left[\,Y_\nu^\dagger\,\mathcal{M}_{\nu}\,
Y_\nu^\ast\,\right]_{11}} {(Y_\nu^{\dag} Y_\nu^{})_{11}}\,,
\label{ep1Mn}
\end{eqnarray}
using the seesaw relation given in Eq.~(\ref{ssaw}).

\section{On the Link between Leptogenesis and Low-Energy CP Violation}
In this section we analyze the possible connection between CP violation at low
energies, measurable for example through neutrino oscillations, and
leptogenesis. Of particular interest are the following questions:
\begin{itemize}
\item{If the strength of CP violation at low energies in neutrino oscillations
is measured, what can one infer about the viability
or non-viability of leptogenesis? In particular, can one have viable leptogenesis
even if there is no CP violation at low energies (i.e.no Dirac and no Majorana phases
at low energies)?}%
\item{From the sign of the BAU, can one predict the sign of the CP asymmetries
at low energies, namely the sign of $\mathcal{J}_{CP}$?}
\end{itemize}
We will show that having an explicit parametrization of $m_D$ (or equivalently
of $Y_\nu=m_D/v$) is crucial not only to determine which phases are responsible
for leptogenesis and which ones are relevant for leptonic CP violation at low
energies, but also to analyze the relationship between these two phenomena.

From the available neutrino oscillation data, one obtains some information on
the effective neutrino mass matrix $\mathcal{M}_\nu$ which can be decomposed in
the following way:
\begin{eqnarray}
U_{\nu}\,d_\nu\,U_{\nu}^{\,T}=\mathcal{M}_{\nu} \equiv
L\,L^T\;,\;L\equiv i\,m_D\,D_R^{-1/2}\,. \label{MnuLL}
\end{eqnarray}
The extraction of $L$ from $\mathcal{M}_\nu$ suffers from an intrinsic
ambiguity \cite{Casas:2001sr} in the sense that, given a particular solution
$L_0$ of Eq.~(\ref{MnuLL}), the matrix $L=L_0\,R$ will also satisfy this
equation, provided that $R$ is an arbitrary orthogonal complex matrix, $R \in
O(3,C)$, i.e. $R\,R^T= 1$. It is useful to take as a reference
solution $L_0 \equiv U_\nu\,d_\nu^{1/2}$, so that:
\begin{eqnarray}
\label{L2} L \equiv U_\nu\,d_\nu^{1/2} R\,.
\end{eqnarray}
Since three of the phases of $m_D$ can be eliminated, the matrix $L$ has 15
independent parameters. The parametrization of $L$ given in Eq.~(\ref{L2}) has
the interesting feature that all its parameters are conveniently distributed
among $U_\nu$, $d_\nu$ and $R$, which contain 6 (3 angles + 3 phases), 3 and 6
(3 angles + 3 phases) independent parameters, respectively. Of the 18
parameters present in the Lagrangian of the fundamental theory described by
$m_D$ and $D_R$, only 9 appear at low energy in $\mathcal{M}_\nu$ through the
seesaw mechanism. To further disentangle $m_D$ from $D_R$ in $L$, one needs the
3 remaining inputs, namely the three heavy Majorana masses of $D_R$. 

Coming back to the connection between leptogenesis and low-energy data, it is
important to note that $U_\nu$ does not appear in the relevant combination for
leptogenesis $Y_\nu^{\dag}Y_\nu^{}$, in the same way as $R$ does not appear in
$\mathcal{M}_\nu$. Indeed, one has:
\begin{eqnarray}
 m_D^\dagger m_D^{} = D_R^{1/2}\,R^{\dagger}\,d_\nu\,R\,D_R^{1/2}
 \,.\label{HHdag}
\end{eqnarray}
From the above discussion, it follows that it is possible to write $m_D$ in the
form $m_D=-i\,U_\nu\,d_{\nu}^{1/2}R\,D_R^{1/2}$ in such a way that leptogenesis
and the low-energy neutrino data (contained in $\mathcal{M}_\nu$) depend on two
independent sets of CP-violating phases, respectively those in $R$ and those in
$U_{\nu}$. In particular, one may have viable leptogenesis even in the limit
where there are no CP-violating phases (neither Dirac nor Majorana) in $U_\nu$
and hence, no CP violation at low energies \cite{Rebelo:2002wj}. Therefore, in
general it is not possible to establish a link between low-energy CP violation
and leptogenesis. This connection is model dependent: it can be drawn only by
specifying a particular \emph{ansatz} for the fundamental parameters of the
seesaw, $m_D$ and $D_R$, as will be done in the following sections.

The relevance of the matrix $R$ for leptogenesis can be rendered even more
explicit \cite{Masina:2002qh} by rewriting the $\varepsilon_1$ asymmetry by
means of Eq.~(\ref{HHdag}) and defining $R_{ij}= |R_{ij}| e^{i
\varphi_{ij}/2}$, $\Delta m^2_\odot \equiv
\Delta m^2_{21} $ and 
$\Delta m^2_@  \equiv  \Delta m^2_{32} $. In the
case of hierarchical heavy Majorana neutrinos, say $M_1 \ll M_2 \ll M_3$ one
obtains
\begin{eqnarray}
\varepsilon_1 \simeq \frac{3}{16\pi} \frac{M_1}{v^2} \frac {
\Delta m^2_@  |R_{31}|^2 \sin \varphi_{31}  - 
\Delta m^2_\odot  |R_{11}|^2
\sin\varphi_{11}} { m_1|R_{11}|^2 +m_2|R_{21}|^2
+m_3|R_{31}|^2}\,, \label{edb}
\end{eqnarray}
and we recover what one would have expected by intuition, namely that the
physical quantities involved in determining $\varepsilon_1$ are just $M_1$, the
spectrum of the light neutrinos, $m_i$, and the first column of $R$, which
expresses the composition of the lightest heavy Majorana neutrino in terms of
the light neutrino masses $m_i$. 

As stressed before, different \emph{ans\"atze} for $R$ have no direct impact on
CP violation at low energy; the impact is in a sense indirect because $R$
specifies if dominance of some heavy Majorana neutrino is at work in the seesaw
mechanism \cite{Lavignac:2002gf}.

In conclusion, the link between leptogenesis and low-energy CP violation can
only be established in the framework of specific \emph{ans\"atze} for the
leptonic mass terms of the Lagrangian. In order to derive a necessary condition 
for such a link to exist, it is convenient to use the following
triangular parametrization for $m_D$ :

\medskip \bigskip
\textbf{Triangular parametrization}
\medskip \bigskip

It can be easily shown that any arbitrary complex matrix can be written as the
product of a unitary matrix U with a lower triangular matrix $Y_{\triangle}$.
In particular, the Dirac neutrino mass matrix can be written as:
\begin{eqnarray}
m_D = v\,U\,Y_{\triangle}\,, \label{mDtri}
\end{eqnarray}
with $Y_{\triangle}$ of the form:
\begin{eqnarray}
Y_{\triangle}= \left(\begin{array}{ccc}
y_{11} & 0 & 0 \\
y_{21}\,e^{i\,\phi_{21}} & y_{22} & 0 \\
y_{31}\,e^{i\,\phi_{31}} & y_{32}\,e^{i\,\phi_{32}} & y_{33}
\end{array}
\right)\,, \label{Ytri1}
\end{eqnarray}
where $y_{ij}$ are real positive numbers. Since $U$ is unitary, in general it
contains six phases. However, three of these phases can be rephased away by a
simultaneous phase transformation on ${\nu}_{L}^{\,0}$, $\ell_{L}^{\,0}$, which
leaves the leptonic charged current invariant. Under this transformation, $m_D
\rightarrow P_{\xi} m_D $, with $P_{\xi}={\rm diag} \left(e^{i \, \xi_1},e^{i
\, \xi_2},e^{i \, \xi_3} \right)$. Furthermore, $Y_{\triangle}$ defined in
Eq.~(\ref{Ytri1}) can be written as:
\begin{eqnarray}
Y_{\triangle}= {P_{\beta}^\dagger}\ {\hat Y_{\triangle}}\ P_{\beta}\,,
\label{Ytri2}
\end{eqnarray}
where $P_\beta ={\rm diag} (1, e^{i \, \beta_1}, e^{i \, \beta_2})$ with
$\beta_1=-\phi_{21}$, $\beta_2=-\phi_{31}$ and
\begin{eqnarray}
{\hat Y_{\triangle}}= \left(\begin{array}{ccc}
y_{11} & 0 & 0 \\
y_{21}  & y_{22} & 0 \\
y_{31}  & y_{32}\,e^{i \, \sigma} & y_{33}
\end{array}
\right) \,,\label{Ytri3}
\end{eqnarray}
with $\sigma=\phi_{32}- \phi_{31}+ \phi_{21}$. It follows then from
Eqs.~(\ref{mDtri})~and~(\ref{Ytri2}) that the matrix $m_D$ can be decomposed in
the form
\begin{eqnarray}
m_D=v\,U_{\rho}\,P_{\alpha}\,{\hat Y_{\triangle}}\,P_{\beta}\,,
\label{mDdec}
\end{eqnarray}
where $P_\alpha ={\rm diag} (1, e^{i \, \alpha_1}, e^{i \, \alpha_2})$ and
$U_\rho$ is a unitary matrix containing only one phase $\rho$. Therefore, in
the WB where $m_\ell$ and $M_R$ are diagonal and real, the phases $\rho$,
$\alpha_1$, $\alpha_2$, $\sigma$, $\beta_{1}$ and $\beta_{2}$ are the only
physical phases characterizing CP violation in the leptonic sector. 

\medskip \bigskip
\textbf{A necessary condition}
\medskip \bigskip

The phases
relevant for leptogenesis are those contained in $m_D^{\,\dagger} m_D^{}$. From
Eqs.~(\ref{Ytri2})-(\ref{mDdec}) we conclude that these phases are $\sigma$,
$\beta_{1}$ and $\beta_{2}$, which are linear combinations of the phases
$\phi_{ij}$. On the other hand, all the six phases of $m_D$ contribute to the
three phases of the effective neutrino mass matrix at low energies
\cite{Branco:2001pq} which in turn controls CP violation in neutrino
oscillations. Since the phases $\alpha_1$, $\alpha_2$ and $\rho$ do not
contribute to leptogenesis, it is clear that 
\textbf{a necessary condition}\cite{Savoy} for a direct
link between leptogenesis and low-energy CP violation to exist is the
requirement that the matrix $U$ in Eq. (\ref{mDtri}) contains no CP-violating
phases. Note that, although the above condition was derived in a specific WB
and using the parametrization of Eq. (\ref{mDtri}), it can be applied to any
model. This is due to the fact that starting from arbitrary leptonic mass
matrices, one can always make WB transformations to render $m_\ell$ and $M_R$
diagonal, while $m_D$ has the form of Eq.~(\ref{mDtri}). A specific class of
models which satisfy the above necessary condition in a trivial way are those
for which $U=1$, leading to $m_D=v\,Y_\triangle$. This condition is
necessary but not sufficient to allow for a prediction of the sign of the CP
asymmetry in neutrino oscillations, given the observed sign of the BAU together
with the low-energy data. More restrictive
class of matrices $m_D$ in triangular form have been considered
and it was shown that, in an appropriate
limiting case, these structures for $m_D$ lead to the ones assumed by Frampton,
Glashow and Yanagida in \cite{FGY}.
Let us consider the following form for $m_D$:
\begin{eqnarray}
m_D=v\,Y_\triangle = v\,\left(\begin{array}{ccc}
 y_{11}   & \; 0       &\; 0 \\
 y_{21}\,e^{i\,\phi_{21}}   &\; y_{22}       &\; 0 \\
y_{31}\,e^{i\,\phi_{31}}   &\; y_{32}\,e^{i\,\phi_{32}} &\; y_{33}
\end{array}\right)\,\label{MDmin}\
\end{eqnarray}
Then, from Eq.~(\ref{epsj1}) the CP asymmetry generated in the decay of the
heavy Majorana neutrino $N_j$ is
\begin{eqnarray}
\label{ejtri} \varepsilon_j=-\frac{1}{8\pi(H_\triangle)_{jj}} \sum_{i\neq
j}\,{\rm Im}  [(H_\triangle)_{ij}^2]\,f_{ij}\,,
\end{eqnarray}
where
\begin{eqnarray}
\label{fij}
 H_\triangle =Y^{\dag}_\triangle Y_\triangle^{} \quad,\quad
 f_{ij}=f\!\left(\frac{M_i^{\,2}}{M_j^{\,2}}\right)\,,
\end{eqnarray}
with $f(x)$ defined in Eq.~(\ref{f}).

From Eqs.~(\ref{MDmin}) and (\ref{fij}) we readily obtain
\begin{eqnarray}
& {\rm Im} [(H_\triangle)_{21}^2]=y_{21}^2\,
   y_{22}^2\sin
(2{{\phi}_{21}}) + 2\,y_{21}\,y_{22}\,y_{31}\,y_{32}\sin
\theta_1\,
    + y_{31}^2\, y_{32}^2\sin \theta_2\,, \nonumber \\ 
&{\rm Im}  [(H_\triangle)_{31}^2]=y_{31}^2\,y_{33}^2\,\sin (2\,{{\phi
}_{31}})\,,  \nonumber \\  
&{\rm Im} [(H_\triangle )_{32}^2]=\,y_{32}^2\,y_{33}^2\sin (2\,{{\phi
}_{32}})\,,\label{imhij}
\end{eqnarray} with
$\theta_1={{\phi }_{21}} + {{\phi }_{31}} - {{\phi }_{32}}$ and
$\theta_2=2\,\left( {{\phi }_{31}} - {{\phi }_{32}} \right)$.

All the information about light neutrino masses and mixing is fully contained in the
effective neutrino mass matrix $\mathcal{M}_\nu$ which is determined through
the seesaw formula given by Eq.~(\ref{ssaw}). In this case
\small
\begin{eqnarray}
\label{Mntri}
\mathcal{M}_\nu=\frac{v^2}{M_1}\left(\begin{array}{ccc} y_{11}^2
&\;y_{11}y_{21}e^{i\,\phi_{21}}\; &\;y_{11}
y_{31}e^{i\,\phi_{31}}\;\\ y_{11}y_{21}e^{i\,\phi_{21}} &y_{21}^2
e^{2i\,\phi_{21}}+y_{22}^2\frac{M_1}{M_2} &y_{21}
y_{31}e^{i(\phi_{31}+\phi_{21})}+y_{22}y_{32}
\frac{M_1}{M_2}\,e^{i\phi_{32}} \\ y_{11} y_{31}e^{i\phi_{31}}
&y_{21}
y_{31}e^{i(\phi_{31}+\phi_{21})}+y_{22}y_{32}\frac{M_1}{M_2}\,
e^{i\phi_{32}}
&y_{31}^2e^{2i\phi_{31}}+y_{33}^2\frac{M_1}{M_3}+y_{32}^2\frac{M_1}{M_2}
\,e^{2i\phi_{32}}
\end{array}\right)
\end{eqnarray}
\normalsize
It follows from Eqs.~(\ref{ejtri})-(\ref{Mntri}) that, in principle, one can
obtain simultaneously viable values for the CP asymmetries $\varepsilon_j$ and
a phenomenologically acceptable effective neutrino mass matrix in order to
reproduce the solar, atmospheric and reactor neutrino data. This can be
achieved by consistently choosing the values of the free parameters $y_{ij}$,
$M_i$ and $\phi_{ij}\ $. Yet, a closer look at Eqs.~(\ref{ejtri})-(\ref{imhij})
shows that there are terms contributing to $\varepsilon_j$ which vanish
independently from the others. This means that a non-vanishing value of
$\varepsilon_j$ can be guaranteed even for simpler structures for $Y_\nu$,
which can be obtained from $Y_\triangle$ assuming additional zero entries in
the lower triangle\footnote{Notice however that the vanishing of diagonal
elements in $Y_\triangle$ would imply ${\rm det}\,(m_D)=0$ and consequently,
${\rm det}\,(\mathcal{M}_\nu)=0$, leading to the existence of massless light
neutrinos.}. A systematic study of the various possible textures, with 
emphasis on the link between low energy data and leptogenesis 
has been done \cite{Savoy}.

\section{Concluding Remarks}
The phenomenon of CP violation plays a crucial role both in Particle Physics 
and Cosmology. In the context of Particle Physics, the breaking of CP 
is closely connected to the least established sectors of the SM, namely
the Higgs sector and the Yukawa sector. The study of CP violation can thus 
be a good ground for the search for New Physics \cite{New} . It is natural to wonder 
whether the various manifestations of CP violation ( at low energies
in the quark and lepton sectors and at high energy in the existence of BAU )
have all a common origin \cite{common}.
We have presented here a brief review of the role of CP violation in
generating BAU through leptogenesis, with emphasis on a possible connection 
between CP violation responsible for leptogenesis and leptonic CP violation   
at low energies, measurable through neutrino oscillations. We have emphasized
that such a connection is only possible within specific ansatze for the flavour
structure of the fundamental leptonic mass matrices. This essentially means 
that the problem of connecting CP violation present in leptogenesis to
CP violation detectable in neutrino oscillations cannot be separable from 
the general flavour problem.

\section*{Acknowledgments}
The author thanks the organizers of  
Flavor Physics And CP Violation (FPCP 2003) 
Paris, France for the stimulating Conference
and for the very warm hospitality.

This work was partially supported by
{\em Funda\c{c}{\~a}o para a Ci{\^e}ncia e a Tecnologia} (FCT,
Portugal) through the projects POCTI/FNU/43793/2002 
(for which part of the funding comes from FEDER, EU)
and CFIF - Plurianual (2/91) and is based on  work done 
in collaboration with R. Gonz\' alez Felipe, F. R. Joaquim, I.
Masina, M. N. Rebelo and C. A. Savoy which had been partially 
supported also by FCT and by the RTN European Program
HPRN-CT-2000-00148.

%%%%%%%%%%%%%%%%%%%%%%%%%%%%%%%%%%%%%
% Label to flag the last page of your contribution
% Replace Perret by your name starting with a capital letter
%
\label{BrancoEnd}
 
\end{document}